\documentclass[12pt,a4paper,final]{iopart}

\usepackage{upgreek}
\usepackage{iopams}  
\usepackage{graphicx}
\usepackage[breaklinks=true,colorlinks=true,linkcolor=blue,urlcolor=blue,citecolor=blue]{hyperref}
\usepackage{float}
\floatstyle{plaintop}
\restylefloat{table}
\usepackage{array}
\newcolumntype{P}[1]{>{\centering\arraybackslash}p{#1}}
\usepackage{makecell}
\usepackage{tabularx}
\expandafter\let\csname equation*\endcsname\relax 
\expandafter\let\csname endequation*\endcsname\relax 
\usepackage{amsmath}
\usepackage{csquotes}

\usepackage[labelfont={bf,footnotesize},textfont=footnotesize,justification=justified,singlelinecheck=false,labelsep=period]{caption}
 
\begin{document}

\newcommand{\mote}{MoTe$_2$}
\newcommand{\wte}{WTe$_2$}
\newcommand{\mowte}{Mo$_{1-\mathrm{x}}$W$_{\mathrm{x}}$Te$_2$}
\newcommand{\td}{T$_d$}
\newcommand{\tp}{1T$^\prime$}
\newcommand{\hp}{2H}
\newcommand{\cm}{cm$^{-1}$}
\newcommand{\degc}{$^{\circ}$C}
\newcommand{\degrees}{$^{\circ}$}
\newcommand\tensor[1]{\overset{\mbox{\tiny$\leftrightarrow$}}{#1}}
\renewcommand{\thefootnote}{\fnsymbol{footnote}}

\def\phasediagram{
        \begin{figure}
          \centering
          \includegraphics[trim = 0in 0in 0in 0in,clip=true,width=86mm]{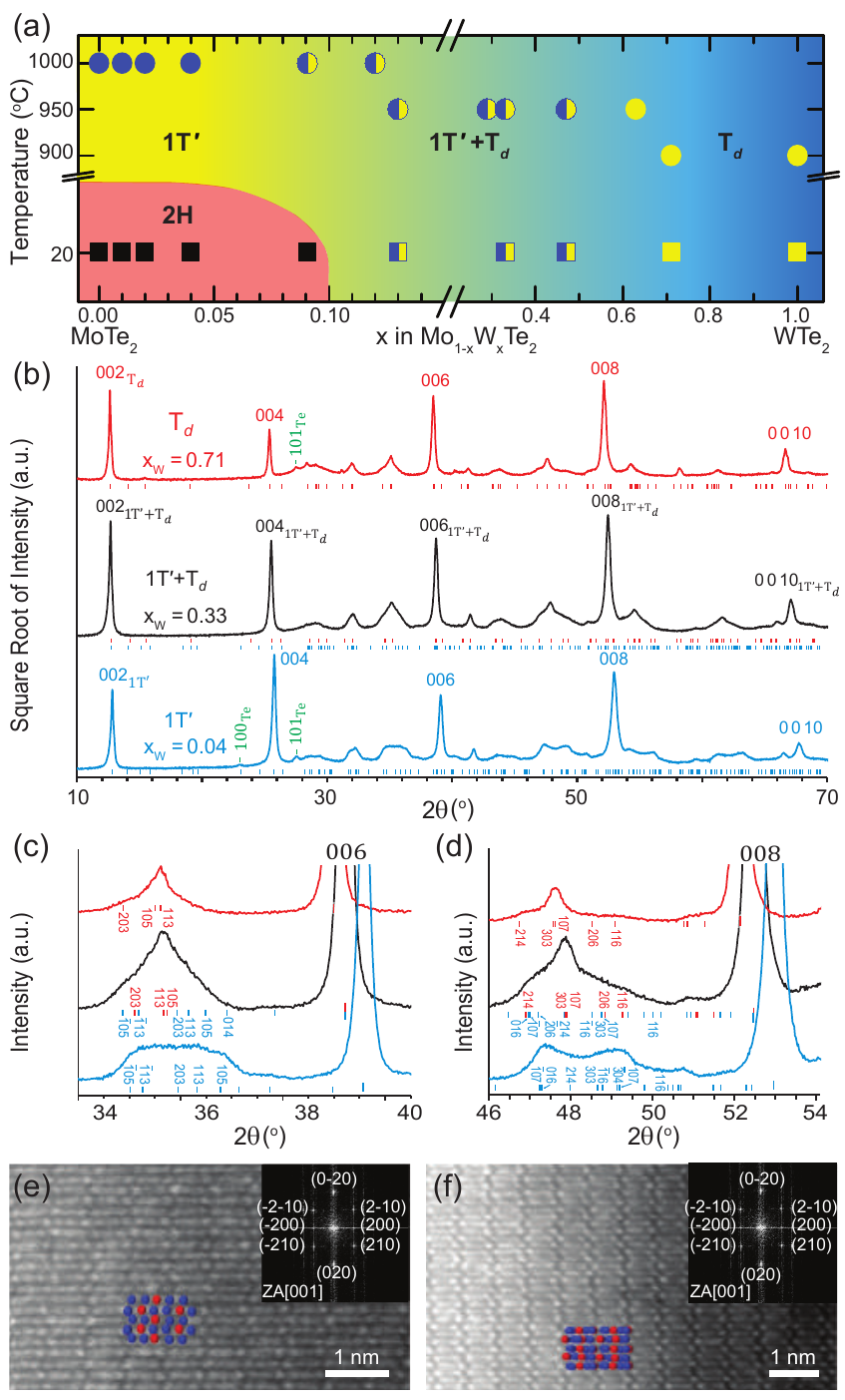}
          \caption{(a) Temperature-composition map of crystal phases in the {\mote}-{\wte} system, identified from the XRD analysis. Circles correspond to the samples that were quenched from the growth temperature indicated on the y-axis (for example, 3 adjacent circles in the upper-left corner represent {\mowte} alloys with x = 0, 0.01, and 0.02 that were quenched from the growth temperature of 1000 {\degc}), and squares correspond to the samples that were additionally annealed in vacuum at 750 {\degc} after growth and slowly cooled to room temperature. Blue circles correspond to {\tp} single-phase alloys, yellow circles and squares correspond to {\td} single-phase alloys, half-filled blue + yellow circles correspond to two-phase {\tp} + {\td} mixture, and black squares are single-phase 2H alloys. The pink area at the bottom-left schematically denotes the single-phase 2H region, which extends from x = 0 to x = 0.09 ${\pm}$ 0.01 at room temperature. (b) XRD $\uptheta$-2$\uptheta$ scans of three most representative {\mowte} samples with x = 0.04 (single-phase {\tp} alloy), x = 0.33 (two-phase {\tp} + {\td} mixture), and x = 0.71 (single-phase {\td} alloy). (c,d) Enlarged XRD scans near 006 and 008 reflections that compare single-phased x = 0.04 ({\tp}) and x = 0.71 ({\td}) alloys with x = 0.33 two-phase mixture. Reflections position and hkl(s) for figure 1(b-d) are listed in table S2 in the Supporting Information. (e,f) HAADF-STEM images of {\mowte} samples with (e) x = 0.04 in {\tp} phase and (f) x = 0.71 in {\td} phase, together with Fast Fourier Transforms (FFT) in the insets. Atomic models are superimposed on the images where Te atoms are blue spheres and Mo/W atoms are red spheres.}
          \label{phasediagram}
        \end{figure}
}

\def\introraman{
        \begin{figure}
          \centering
          \includegraphics[trim = 0in 0in 0in 0in,clip=true,width=6.2in]{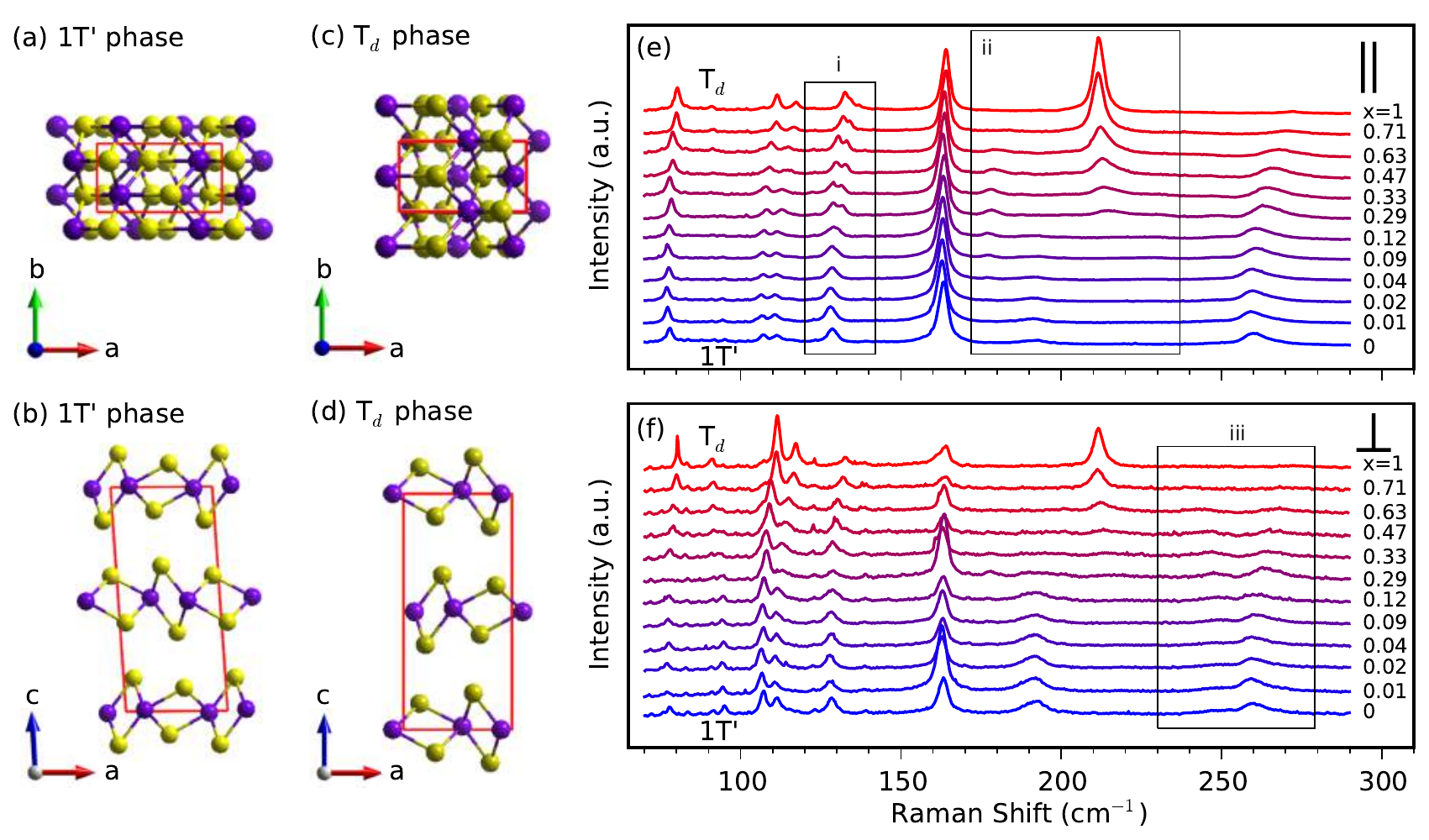}
          \caption{(a,b) Three layer {\tp}-{\mote} and (c,d) {\td}-{\wte} crystal structure diagrams. The red boxes indicate the unit cells. (e,f) Polarized Raman spectra for {\mowte} alloys with all crystal orientations summed for co-polarized ($\parallel$) and cross-polarized ($\perp$) configurations, respectively. The x value for each alloy is labeled. Boxed regions are shown in figure~\ref{compdep}.}
          \label{introraman}
        \end{figure}
}

\def\compdep{
        \begin{figure}
          \centering
          \includegraphics[trim = 0in 0in 0in 0in,clip=true,width=6.2in]{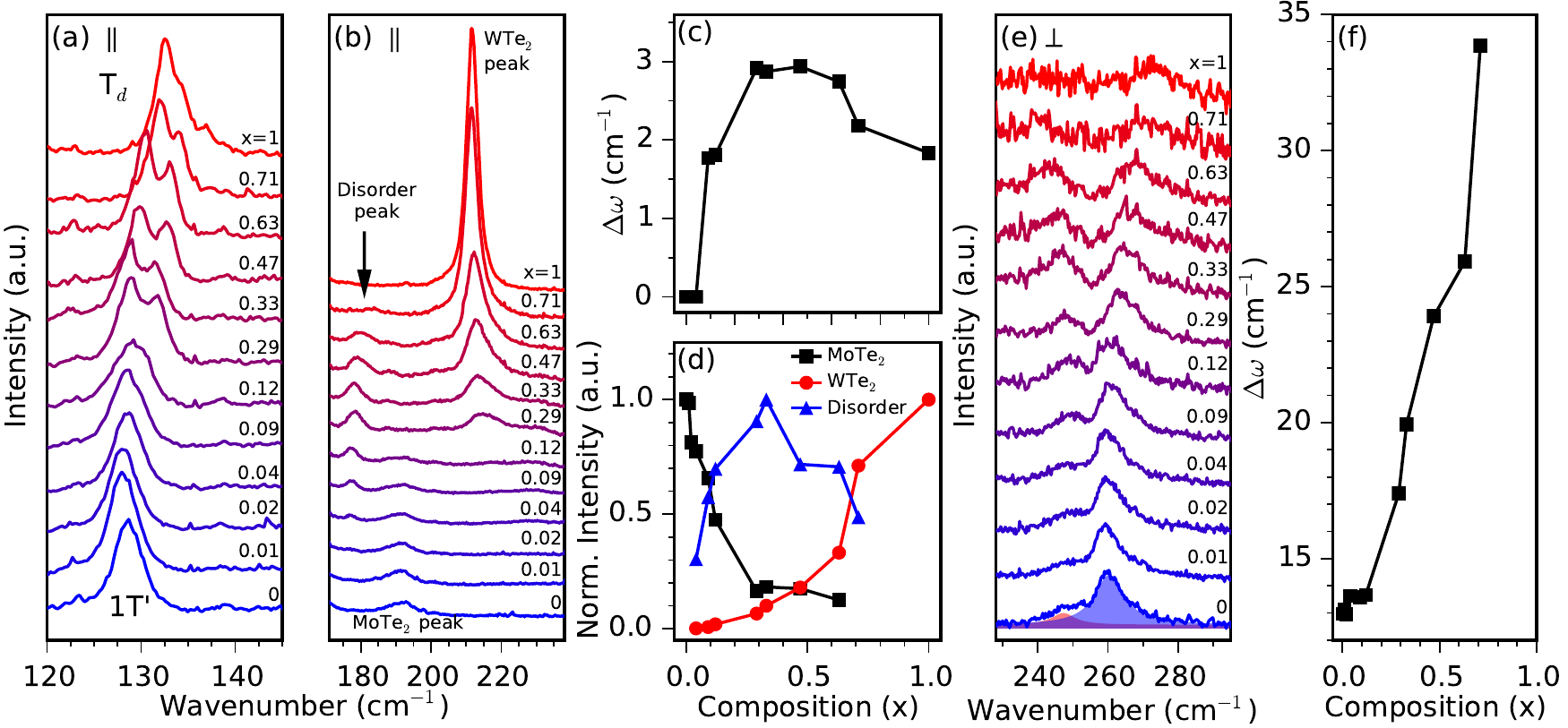}
          \caption{(a) Magnified Raman spectra from box i of figure~\ref{introraman}(e) highlighting the splitting of the 128 {\cm} peak. (b) Magnified Raman spectra from box ii of figure~\ref{introraman}(e). The {\tp}-{\mote} mode at 192 {\cm} decreases with x while the {\td}-{\wte} mode at 212 {\cm} increases. A new ``disorder peak'' exclusive to the alloys appears at 178 {\cm}. (c) Separation between the doublet in (a) versus x. (d) Normalized intensity versus x for the 192 {\cm} {\tp}-{\mote} mode (black squares), 212 {\cm} {\wte} mode (red circles), and the 178 {\cm} mode (blue triangles) identified in (b). (e) Magnified Raman spectra from box iii of figure~\ref{introraman}(f). Example fits are shown as shaded blue and red Lorentzian functions. (f) The peak separation of the doublet in (e), $\Delta \omega$, versus x extracted from fits to the data in (e).}
          \label{compdep}
        \end{figure}
}

\def\hraman{
        \begin{figure}
          \centering
          \includegraphics[trim = 0in 0in 0in 0in,clip=true,width=86mm]{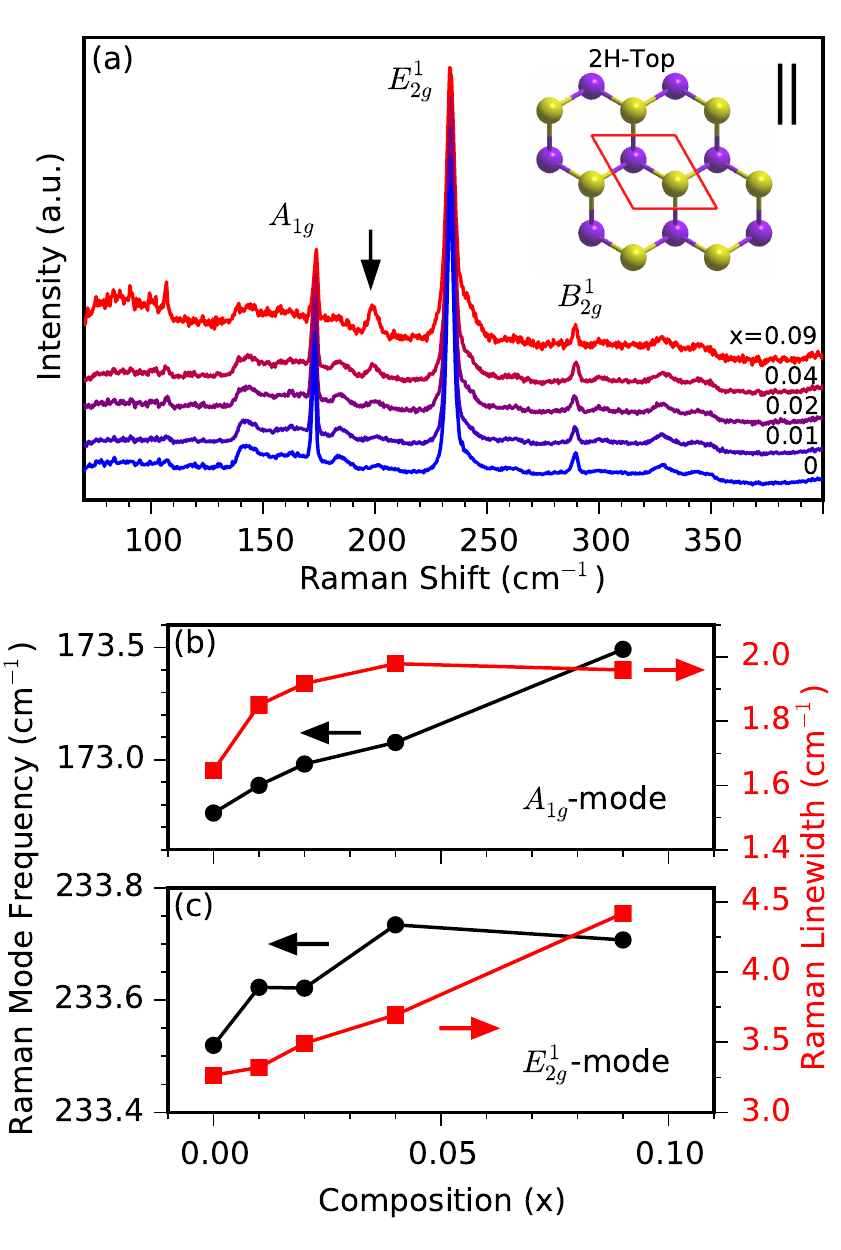}
          \caption{(a) Normalized Raman spectra of 2H-{\mowte} alloys measured in the co-polarized configuration. The double-resonance Raman mode is indicated with the black arrow. Inset: crystal structure of 2H-{\mote}. (b) and (c) illustrate the evolution in frequencies (black, left axis) and linewidths (red, right axis) of the $A_{1g}$ and $E_{2g}^1$ modes, respectively.}
          \label{hraman}
        \end{figure}
}

\def\scm{
        \begin{figure}
          \centering
          \includegraphics[trim = 0in 0in 0in 0in,clip=true,width=86mm]{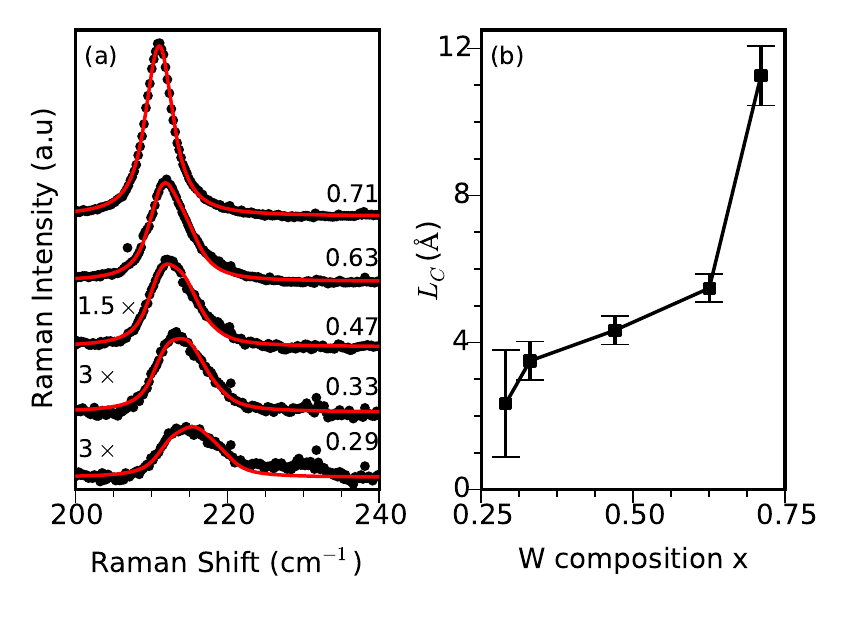}
          \caption{(a) Composition-dependence of the 212 {\cm} {\td}-{\wte} mode. The red lines are fits to the phonon confinement model in Equation \eqref{pcm}. (b) $L_C$ extracted from the fits in (a) versus composition.}
          \label{scm}
        \end{figure}
}

\def\modetable{
        \renewcommand{\arraystretch}{0.53}
        \begin{table}
            \begin{center}
              \begin{tabular}{ |P{1.25cm} P{1.25cm} P{1.75cm}|P{1.25cm} P{1.75cm}|P{1.25cm} P{1.25cm} P{1.75cm}| }
                  \hline
                  \multicolumn{3}{|c|}{{\tp}-{\mote}} & \multicolumn{2}{c|}{{\td}-{\mote}} & \multicolumn{3}{c|}{{\td}-{\wte}} \\
                  \hline
                  
                  \multicolumn{1}{|c}{\thead{$\omega_{calc}$ \\ ({\cm})}} & \multicolumn{1}{|c}{\thead{$\omega_{exp}$ \\ ({\cm})}} & \multicolumn{1}{|c|}{Symmetry} & \multicolumn{1}{c}{\thead{$\omega_{calc}$ \\ ({\cm})}}  & \multicolumn{1}{|c|}{Symmetry} & \multicolumn{1}{c}{\thead{$\omega_{calc}$ \\ ({\cm})}}  & \multicolumn{1}{|c}{\thead{$\omega_{exp}$ \\ ({\cm})}} & \multicolumn{1}{|c|}{Symmetry}\\
                  \hline
                  76.81 & 78 & $A_g$ & 76.96 & $A_1$ & 79.0 & 80 & $A_1$\\
                  85.56 & & $A_g$ & 85.74 & $B_1$ & 84.6 & 91 &$A_2$\\
                  88.20 & & $B_g$ & 88.09 & $B_2$ & 84.7 & & $B_2$\\
                  90.88 & & $B_g$ & 90.68 & $A_2$ & 86.2 & & $B_1$\\
                  104.90 & 107 & $B_g$ & 104.88 & $A_2$ & 107.7 & & $B_2$\\
                  105.61 & 111 & $B_g$ & 105.52 & $B_2$ & 107.7 & & $B_2$\\
                  108.37 & & $A_u$ & 108.37 & $A_2$ & 110.9 & 111 &$A_2$\\
                  108.67 & & $A_g$ & 108.71 & $A_1$ & 111.5 & 117 & $A_1$\\
                  110.80 & & $A_u$ & 110.76 & $B_2$ & 112.6 & & $B_2$\\
                  113.60 & 111 & $A_g$ & 113.61 & $B_1$ & 115.3 & &$B_1$\\
                  115.38 & & $B_u$ & 115.53 & $B_1$ & 120.6 & & $B_1$\\
                  123.82 & & $B_u$ & 123.23 & $A_1$& 127.2 & 132 & $A_1$\\
                  125.52 & & $A_g$ & 126.23 & $A_1$ & 127.4 & & $B_1$\\
                  128.25 & 128 & $A_g$ & 128.07 & $B_1$ & 128.5 & 134 & $A_1$\\
                  129.60 & & $B_u$ & 129.92 & $B_1$ & 130.4 & 137 & $A_1$\\
                  134.80 & & $B_u$ & 134.84 & $A_1$ & 131.3 & & $B_1$\\
                  155.54 & & $A_g$ & 155.59 & $B_1$ & 146.3 & & $A_2$\\
                  159.24 &163&$A_g$ & 159.35 & $A_1$ & 146.6 & & $A_2$\\
                  176.52 & & $A_u$ & 176.64 & $A_2$ & 152.4 & & $B_1$\\
                  176.97 & & $A_u$ & 176.87 & $B_2$ & 153.7 & & $A_2$\\
                  187.99 & & $B_g$ & 188.27 & $A_2$ & 155.4 & & $A_1$\\
                  189.20 & 192 & $B_g$& 189.29 & $B_2$ & 156.2 & & $B_2$\\
                  192.13 & & $B_u$ & 192.11 & $B_1$ & 165.6 & & $B_1$\\
                  192.33 & & $B_u$ & 192.25 & $A_1$ & 166.2 & 164 & $A_1$\\
                  247.13 & & $A_g$ & 247.24 & $A_1$ & 201.2 & & $A_1$\\
                  249.03 & & $A_g$ & 248.99 & $B_1$ & 201.5 & & $B_1$\\
                  251.58 & 251 & $A_g$ & 251.48 & $A_1$ & 204.6 & 212 & $A_1$\\
                  254.14 & 260 & $A_g$ & 253.95 & $B_1$ & 206.1 & & $B_1$\\
                  265.96 & & $B_u$ & 266.14 & $B_1$ & 227.9 & & $A_1$\\
                  267.37 & & $B_u$ & 267.42 & $A_1$ & 228.4 & & $B_1$\\
                  \hline
              \end{tabular}
            \end{center}
        \caption{Theoretically predicted and experimentally measured wavenumbers, $\omega_{calc}$ and $\omega_{exp}$, respectively, alongside their associated symmetries.}
        \label{modetable}
        \end{table}
}

\title[Evolution of Raman spectra in {\mowte} alloys]{Evolution of Raman spectra in {\mowte} alloys}

\author{Sean M. Oliver$^{1,*}$, Ryan Beams$^{2,*}$, Sergiy Krylyuk$^{2,3}$, Irina Kalish$^2$, Arunima K. Singh$^2$, Alina Bruma$^2$, Francesca Tavazza$^2$, Jaydeep Joshi$^1$, Iris R. Stone$^1$, Stephan J. Stranick$^2$, Albert V. Davydov$^2$, Patrick M. Vora$^{1,\dagger}$}

\address{$^1$Department of Physics and Astronomy, George Mason University, Fairfax, VA 22030 USA}
\address{$^2$Material Measurement Laboratory, National Institute of Standards and Technology, Gaithersburg, MD 20899, USA}
\address{$^3$Theiss Research, La Jolla, CA 92037, USA}
\address{$^*$These authors contributed equally to the work.}
\address{$^\dagger$Author to whom correspondence should be addressed, \mailto{pvora@gmu.edu}.}

\begin{abstract}
\noindent{}The structural polymorphism in transition metal dichalcogenides (TMDs) provides exciting opportunities for developing advanced electronics. For example, {\mote} crystallizes in the {\hp} semiconducting phase at ambient temperature and pressure, but transitions into the {\tp} semimetallic phase at high temperatures. Alloying {\mote} with {\wte} reduces the energy barrier between these two phases, while also allowing access to the {\td} Weyl semimetal phase. The {\mowte} alloy system is therefore promising for developing phase change memory technology. However, achieving this goal necessitates a detailed understanding of the phase composition in the {\mote}-{\wte} system. We combine polarization-resolved Raman spectroscopy with X-ray diffraction (XRD) and scanning transmission electron microscopy (STEM) to study {\mowte} alloys over the full compositional range x from 0 to 1. We identify Raman and XRD signatures characteristic of the {\hp}, {\tp}, and {\td} structural phases that agree with density-functional theory (DFT) calculations, and use them to identify phase fields in the {\mote}-{\wte} system, including single-phase {\hp}, {\tp}, and {\td} regions, as well as a two-phase {\tp} + {\td} region. Disorder arising from compositional fluctuations in {\mowte} alloys breaks inversion and translational symmetry, leading to the activation of an infrared {\tp}-{\mote} mode and the enhancement of a double-resonance Raman process in {\hp}-{\mowte} alloys. Compositional fluctuations limit the phonon correlation length, which we estimate by fitting the observed asymmetric Raman lineshapes with a phonon confinement model. These observations reveal the important role of disorder in {\mowte} alloys, clarify the structural phase boundaries, and provide a foundation for future explorations of phase transitions and electronic phenomena in this system.
\end{abstract}

\vspace{2pc}
\noindent{}{\it Keywords}: Transition metal dichalcogenides, alloys, Raman spectroscopy, phase transition, disorder, X-ray diffraction, polarization\\

\maketitle

\section{Introduction}
Transition metal dichalcogenides (TMDs) are van der Waals (vdW) compounds that follow the general formula of MX$_2$, where M is a transition metal from Groups IVB-VIB, and X is a Group VIIA chalcogen, such as S, Se, and Te. This chemical versatility leads to unique electronic properties, such as semiconducting behavior \cite{Mak2016}, superconductivity \cite{Qi2015, Pan2015, Kang2015}, and topological electronic states \cite{Qian2014, Soluyanov2015, Xu2016, Tamai2016}. Furthermore, two-dimensional (2D) TMD monolayers can be vertically stacked without the need for lattice matching, which allows these dissimilar electronic phases to be combined in a single heterostructure \cite{Geim2013}.

The chemical versatility intrinsic to TMDs and the novel interactions achieved through vdW stacking are complimented further by the structural polymorphism of TMDs. A prototypical example is molybdenum ditelluride ({\mote}), which can be grown in a semiconducting {\hp} phase (space group P6$_3$/mmc) or a semimetallic \tp{} phase (space group P2$_1$/m) \cite{Park2015, Wilson1969, Dawson1987}. The hexagonal {\hp} phase, characterized by a trigonal prismatic coordination, is thermodynamically stable under ambient conditions, while the monoclinic {\tp} phase is stable above 900 \degc{}. The {\tp} phase can be stabilized at room temperature by rapid cooling \cite{Keum2015}, control of the tellurization rate of Mo films \cite{Park2015}, or choosing appropriate precursors for chemical vapor deposition \cite{Naylor2016, Zhou2015b}. When cooled below $\sim$250 K, {\tp}-{\mote} transitions into an orthorhombic {\td} phase (space group Pmn2$_1$) with broken inversion symmetry as evidenced by electrical, structural, and optical measurements \cite{Wilson1969, Hughes1978, Clarke1978, Chen2016a}.

Interest in {\mote} has surged due to the unique electronic properties of its structural phases as well as the possibility of engineering controlled transitions between these phases. For instance, type-II Weyl semimetal states were theoretically predicted and experimentally observed in both {\td}-{\mote} and {\td}-{\wte} \cite{Soluyanov2015, Xu2016, Sun2015, Chang2016, Wang2016, Huang2016a, Deng2016}. The broken inversion symmetry of the {\td} phase is a necessary condition for the type-II Weyl state \cite{Soluyanov2015, Sun2015, Chang2016} and allows for fundamental studies of interesting topological physics. However, efforts to directly observe the Weyl state using angle-resolved photoemission spectroscopy are frustrated by the presence of overlapping band-crossings and insufficient experimental resolution \cite{Chang2016}. A more practical application driving investigations of {\mote} is the development of atomically thin phase change memory. {\mote} has a small energy difference between the {\hp} and {\tp} phases, making the prospect of engineering on-demand transitions with low power consumption realistic \cite{Duerloo2014,Duerloo2015a,Li2016,Zhang2016a}. Unfortunately, reversible and rapid phase changes in {\mote} have yet to be demonstrated. This may suggest that the energy difference between the {\hp} and {\tp} phases must be reduced further in order to successfully perform phase change operations.

The limitations of {\mote} highlighted above can be addressed by alloy engineering. Substitutional doping of Mo with W atoms results in {\mowte} alloys which have properties advantageous for both fundamental investigations and practical applications. {\mowte} alloys have been theoretically predicted \cite{Chang2016} and experimentally confirmed \cite{Belopolski2016} to be type-II Weyl semimetals. Importantly, the separation of the Weyl nodes in the alloys can be tuned with composition \cite{Chang2016}, which facilitates the observation of topological electronic states. Additionally, the ground state energy difference between the semiconducting {\hp} and semimetallic {\tp} or {\td} phases in {\mowte} alloys can also be tuned with composition \cite{Duerloo2015a,Zhang2016a}, thereby reducing the energy required to trigger a semiconductor-semimetal phase transformation. The desirable combination of tunable phase transitions with the low-dimensionality of TMDs makes {\mowte} highly promising for phase change memory applications.

Further application of the {\mowte} alloy system necessitates an understanding of the compositional dependence of phase transformations and the impact of disorder upon the material properties. The literature on this subject is very limited. In a pioneering work on {\mowte} polycrystalline powders, Revolinsky and Beerntsen found that the alloys crystallize in the {\hp} phase for x $\leq$ 0.15, the {\td} phase for x $\geq$ 0.65, and a two-phase region of {\hp} + {\tp} in between \cite{Revolinsky1964}. On the contrary, Champion detected a two-phase {\hp} + {\td} region only for x = 0.25 composition, whereas a higher (lower) W content resulted in {\mowte} powders in a {\td} ({\hp}) phase, respectively \cite{CHAMPION1965}. However, no detailed structural studies were reported in these two papers to shed light on the co-existence of the {\hp} and {\td} phases, especially considering a noticeable difference in their symmetry.  Recently, Rhodes et al. reported a simplified phase diagram without two-phase regions for single-crystalline {\mowte} alloys grown by the chemical vapor transport method (CVT) \cite{Rhodes2016}. Finally, Lv et al. suggested that a mixed {\tp} + {\td} region exists for 0.5 $<$ x $<$ 0.7 based on Raman measurements \cite{Lv2017}. This significant disagreement between the studies about phase boundaries between the phases in the {\mote}-{\wte} system, combined with the absence of comprehensive studies of compositional disorder on optical properties, calls for a fresh look at this alloy system. 

Here, we address these disagreements in the literature by combining X-ray diffraction (XRD), scanning transmission electron microscopy (STEM), density-functional theory (DFT), and polarization-resolved Raman spectroscopy to explore the properties of the {\tp}, {\td}, and {\hp} structural phases in the {\mowte} crystals grown by iodine-assisted CVT. XRD and STEM measurements indicate that the {\mowte} alloys with the {\tp} crystal structure are stable at elevated temperatures with W content x $\leq$ 0.04, while alloys with the {\td} structure are stable for x $\geq$ 0.63. The alloys with intermediate compositions 0.04 $<$ x $<$ 0.63 form a two-phase, {\tp} + {\td}, mixture. Polarized Raman measurements offer further insight into the transition from {\tp} to the two-phase, {\tp} + {\td}, field and ultimately to the {\td} single-phase region as a function of x. We use Raman tensor analysis to assign the phonon mode symmetry for all compositions and find that the tensor elements must be complex in order to capture the polarization dependence. This observation is consistent with prior studies of {\mote} \cite{Chen2016a,Beams2016} as well as studies of other layered TMD materials \cite{Ribeiro2015}. The Raman peaks for certain phonon modes show particular sensitivity to x and lattice symmetry. For example, the Raman peak at 128 {\cm} for the {\tp}-{\mowte} alloys broadens at $\mbox{x}=0.09$ and splits into a doublet for compositions x $\geq$ 0.29, which implies a loss of inversion symmetry \cite{Chen2016a} due to the substitution of Mo by W. We also observe the activation of a new Raman mode at 178 \cm{} that is unique to $0.02 < \mbox{x} < 1$ alloys. Based on our DFT calculations, we assign this feature as a disorder-activated infrared mode in {\mote}. Furthermore, the separation between the two modes near 260 \cm{} in {\mote} is highly composition-dependent and can be used to infer x. In {\hp}-\mowte{}, we observe minor changes in the $A_{1g}$ and $E_{2g}^1$ mode frequencies, linewidths, and relative intensities. We also identify a new Raman mode at 202~\cm{} that originates from a double-resonance Raman process \cite{Guo2015} and is apparently enhanced by alloy disorder. The comprehensive structural and spectroscopic data assembled here provide an important roadmap for the future study and application of {\mowte} alloys.

\section{Methods}
Polycrystalline {\mowte} alloys (x $= 0...1$) served as precursors for synthesis of single crystals, and were prepared by reacting stoichiometric amounts of molybdenum (99.999 \%), tungsten (99.9 \%), and tellurium (99.9 \%) powders at 750~\degc{} in vacuum-sealed quartz ampoules. {\mowte} crystals were then grown by the CVT method using approximately 1.5 g of poly-{\mowte} charge and a small amount of iodine (99.8 \%, 5 mg/cm$^3$) sealed in evacuated quartz ampoules. It was found that the temperature required for high-yield synthesis of {\mowte} crystals is lower for higher x. Therefore, the growth temperatures used in this study were 1000~\degc{} for x $\leq$ 0.12, 950~\degc{} for 0.12 $<$ x $\leq$ 0.63, and 900~\degc{} for x $\geq$ 0.71. The ampoules were ice-water quenched after 7 days of growth. To study phase transformation in {\mowte}, as-grown crystals were vacuum-sealed in small ampoules (internal volume $\approx 1$ cm$^3$) and annealed at 750~\degc{} for 72 hours followed by cooling to room temperature at a rate of 10~\degc{}/hour.

Chemical compositions with an accuracy of ±0.01 were determined by energy-dispersive X-ray spectroscopy (EDS) using a JEOL JSM-7100F field emission scanning electron microscope (FESEM) equipped with an Oxford Instruments X-Max 80 EDS detector.\footnote[3]{Disclaimer: Certain commercial equipment, instruments, or materials are identified in this paper in order to specify the experimental procedure adequately. Such identification is not intended to imply recommendation or endorsement by the National Institute of Standards and Technology, nor is it intended to imply that the materials or equipment identified are necessarily the best available for the purpose.} We examined the $\uptheta$-$2\uptheta$ XRD patterns derived from a Norelco Philips Diffractometer with the Bragg-Brentano geometry. Lattice parameters were refined using the MDI-JADE 6.5 software package. For the powder XRD study, {\mowte} crystals were finely ground using an agate mortar. An Aberration-Corrected High Angle Annular Dark Field Scanning Transmission Electron Microscopy (Cs-corrected HAADF-STEM) FEI Titan 80-300 TEM/STEM operating at 300 kV was employed for the characterization of {\mowte} samples. The flakes were crushed in ethanol and a drop of solution was deposited onto an amorphous Carbon (a-C) coated TEM grid (Agar Inc.). HAADF-STEM images were collected at a camera length of 100 mm corresponding to inner and outer collection angles of 70.6 and 399.5 mrad respectively.

For Raman measurements, the as-grown {\mowte} crystals were mechanically exfoliated onto Si/SiO$_{\mbox{2}}$ substrates. Polarization-dependent Raman measurements were performed on bulk flakes in a back-scattering geometry at room temperature in atmosphere using a linearly-polarized 532 nm continuous wave laser. The polarization of the excitation beam was controlled with a motorized achromatic half-wave plate and was focused onto the sample using a 0.75 NA microscope objective. The back-scattered Raman emission was collected by the same objective, and then sent through a motorized analyzer and a long-pass filter. The excitation polarization and collection analyzer were oriented in both co-polarized ($\parallel$) and cross-polarized ($\perp$) configurations, and then rotated together while the sample remained fixed. The filtered light was focused into a multimode fiber to scramble polarization and then directed to a spectrometer for analysis. Raman peaks were fit to Lorentzian functions to extract mode frequencies, linewidths, and amplitudes. For angle-dependent Raman maps, measurements were normalized by the feature with the greatest intensity, the $\sim$163 \cm{} peak. The angle-dependent peak intensities were fit using the Raman tensors to assign peak symmetries.

All simulations were based on density-functional theory (DFT) using the projector-augmented wave method as implemented in the plane-wave code VASP \cite{Kresse1996}. The simulations were performed using the vdW-DF-optB88 exchange-correlation functional \cite{Klimes2011}, which provides an excellent description of the lattice constants of bulk {\tp}-{\mote}, {\td}-{\mote}, and {\td}-{\wte}. An energy cutoff of 600 eV and k-point  mesh of 10 $\times$ 18 $\times$ 5 for the 1$\times$1$\times$1 unit cells of bulk {\tp}-{\mote}, {\td}-{\mote}, and {\td}-{\wte} resulted in an accuracy of the total energies of 1 meV/unit cell.  The 5s$^2$5p$^4$ and 4d$^5$5s electrons were considered as the valence  electrons for Te and Mo, respectively. Including the semicore 4s$^2$p$^6$ electrons for Mo had a negligible effect on the results, as, for instance, the lattice parameters of bulk {\tp}-{\mote} changed by $< 0.10$. $\Gamma$-point phonon frequencies of bulk {\tp}-{\mote}, {\td}-{\mote}, and {\td}-{\wte} were estimated from density-functional perturbation theory simulations of the 1$\times$1$\times$1 unit cells of respective materials. Irreducible  representations of normal modes were obtained from the PHONOPY program \cite{Togo2015a} and the Bilbao Crystallographic Server \cite{Aroyo2006}. The phonon dispersion of {\td}-{\wte} in the entire Brillouin zone was estimated by computing normal mode frequencies on a uniform three-dimensional mesh of 51$\times$51$\times$51 $\vec{q}$-points between (0, 0, 0) and ($2\pi/a$, $2\pi/b$, $2\pi/c$) (figure S1 in the Supporting Information). The phonon dispersion of {\td}-{\wte} in the entire Brillouin zone was computed using the finite difference method on the 106 atom 3 $\times$ 3 $\times$ 1 supercell.

\section{Results and Discussion}
\phasediagram{}

Figure~\ref{phasediagram}(a) summarizes heat-treatment schedules, compositions, and crystal phases of {\mowte} (x = 0...1) samples examined in this study. The high-temperature phases of the alloys were preserved by quenching of the growth ampoules in an ice-water bath. This process is known to prevent reversal of the {\tp} phase to the {\hp} phase, which is thermodynamically stable in {\mote} under ambient conditions \cite{Revolinsky1966}. Notably, XRD $\uptheta$-2$\uptheta$ scans from the as-grown, un-milled {\mowte} flakes produce only 00$l$-type reflections and miss all asymmetric reflections, thus limiting the ability to reliably determine phase composition in the alloys. Therefore, we collected the scans from finely ground flakes to register all possible $hkl$ reflections to distinguish {\tp}, {\td}, and {\hp} phases and their mixtures. For example, R. Clarke et al. have established that a {\tp} $\rightarrow$ {\td} transition in {\mote} and the two-phase region can be observed by specifically monitoring $h$0$l$ reflections as a function of temperature, where $\bar{1}$ 0 12 and 1 0 12 reflections of {\tp}-{\mote} coalesce into a single 1 0 12 reflection of the low-temperature {\td} phase \cite{Clarke1978}. A similar approach was applied to construct a phase diagram of Mo$_{1-\mathrm{x}}$Nb$_\mathrm{x}$Te$_2$ alloys that undergo an orthorhombic to monoclinic phase transition with increasing x \cite{Sakai2016}. The $\uptheta$-2$\uptheta$ scans of three representative {\mowte} samples with x = 0.04, 0.33, and 0.71, produced by milling as-grown flakes in an agate mortar, are shown in figure~\ref{phasediagram}(b). Figures~\ref{phasediagram}(c-d) show enlarged portions of the scans around 2$\uptheta$ angles of 35\degrees and 48\degrees, respectively, which illustrate the distinct changes in the lineshape with increasing x. The scans for x = 0.04 and 0.71 were unambiguously assigned to the {\tp} and {\td} phase, respectively. The x = 0.33 scan can only be fitted by combining reflections from both {\tp} and {\td} phases, which indicates a two-phase coexistence. Calculated lattice parameters and Bragg reflection angles for the three samples are presented in tables S1 and S2 in the Supporting Information. By analysis of the powder XRD scans, we established that the quenched {\mowte} alloy samples synthesized in this study are in the monoclinic {\tp} phase for x $\leq$ 0.04, the orthorhombic {\td} phase for x $\geq$ 0.63, and in the {\tp} + {\td} two-phase state for the compositions x between 0.04 and 0.63.

We further verified these observations by performing HAADF-STEM measurements of {\tp}-Mo$_{0.96}$W$_{0.04}$Te$_2$ and {\td}-Mo$_{0.29}$W$_{0.71}$Te$_2$ crystals, shown in figures~\ref{phasediagram}(e, f) with the overlapped atomic models and their corresponding Fast Fourier Transforms (FFT) in the insets.  Both the {\tp} and the {\td} phases exhibit a \enquote{buckled} structure with visible shifts for Te atoms and a zig-zag pattern for Mo/W atoms. The presence of the two phases was observed on a sample with x = 0.33, proving the {\tp} + {\td} coexistence in {\mowte} alloys for 0.04 $<$ x $<$ 0.63, although we were not able to map the spatial distribution of {\tp} and {\td} phases.

In order to study temperature-induced phase transformations in {\mowte} alloys, the samples were annealed in vacuum-sealed ampoules at 750 {\degc} for 72 h followed by slow cooling to room temperature (squares in figure~\ref{phasediagram}(a)). We found that alloys with x $\leq$ 0.09 $\pm$ 0.01 could be converted to the hexagonal {\hp} phase, as schematically depicted in figure~\ref{phasediagram}(a) by the pink-colored area. HAADF-STEM and XRD data of the Mo$_{0.91}$W$_{0.09}$Te$_2$ sample converted from {\tp} into {\hp} phase are provided in figure S2 of the Supporting Information. The vacuum annealing did not change the crystal structures of the alloys with larger x. Thus, an upper limit for {\mowte} alloys to experience a reversible phase transformation between semiconducting {\hp} and metallic {\tp} phases is x $\approx$ 0.09. Approximately the same boundary between metallic and semiconducting phases was recently reported for {\mowte} alloys grown in the 750 {\degc} –- 850 {\degc} temperature range, i.e., below the {\tp} phase existence in bulk {\mote} \cite{Rhodes2016,Lv2017}.

\introraman{}

We now investigate the impact of composition, disorder, and crystal structure on the Raman-active phonon modes of {\mowte} alloys. We first examine the \tp{} $\rightarrow$ \td{} phase transition in {\mowte}, and then explore the impact of alloy potential fluctuations on the {\hp} phase. Due to possible surface oxidation of TMD layers in air \cite{Liu2016b} and to minimize the oxide layer contribution to the detected Raman signal, we focus exclusively on bulk {\mowte} flakes that were mechanically exfoliated onto Si substrates with a 285 nm SiO$_2$ layer. The home-built confocal Raman microscope used in these measurements is oriented in a backscattering geometry and operated in two polarization configurations: one with the excitation polarization and analyzer co-polarized ($\parallel$) and the other with them cross-polarized ($\perp$). The excitation/analyzer orientation is fixed and the two are rotated together relative to the crystal lattice. All measurements are performed at room temperature in atmosphere on bulk flakes. By acquiring a series of these spectra at different orientations, we assemble polarized Raman maps that provide a concise visualization of the angle-dependent Raman spectra as a function of x, which are shown for all {\tp} and {\td} samples in figure S3 of the Supporting Information. Due to our experimental geometry, only $A_g$ ($A_1$) and $B_g$ ($A_2$) symmetry modes are accessible for the {\tp} ({\td}) crystal structure. These modes have distinct dependencies on laser-analyzer orientation, and the orientation relative to the crystal axes \cite{Beams2016}. In short, the polarized Raman signal is given by $I(\theta)=|\hat{e}_s \cdot \tensor{R} \hat{e}_i|^2$, where $\hat{e}_i$ and $\hat{e}_s$ are the incident and scattered fields and $\tensor{R}$ is the Raman tensor. In bulk {\mote} and {\wte}, $\tensor{R}$ is complex-valued for all modes, suggesting that optical absorption is significant \cite{Beams2016,Jiang2016a,Song2016}. We summarize the results of the Raman tensor analysis fitting to the Raman peaks of {\tp}-{\mote} and {\wte} in figures S4 and S5 of the Supporting Information, respectively, and in Table~\ref{modetable} we summarize the experimentally-determined mode assignments for {\tp}-{\mote} and {\td}-{\wte} as well as the results of our DFT calculations of the {\tp}-{\mote}, {\td}-{\mote}, and {\td}-{\wte}. In the {\tp} phase, the $A_g$ and $B_g$ modes are Raman-active while the $A_u$ and $B_u$ modes are only infrared-active. Interestingly, all modes are Raman active for the {\td} phase.

\modetable{}

The polarized Raman maps presented in figure S3 of the Supporting Information are instructional for an overview of the composition-induced evolution of the vibrational modes from the {\tp} phase to the {\td} phase. However, the sensitivity of the Raman spectrum to the orientation of the excitation and analyzer complicates further interpretation of the data in this form. To eliminate this orientation-dependence, we sum the parallel and perpendicular data over all angular orientations (figures~\ref{introraman}(e,f)). The composition-dependent Raman spectra of the alloys show several important features. Most modes are present in all compositions and exhibit only small frequency shifts due to the similarity between the {\tp} and {\td} lattices (figures~\ref{introraman}(a-d)). However, certain Raman modes (identified with boxes in figures \ref{introraman}(e,f)) exhibit unique behavior that is dependent on structural symmetry and composition x in the alloys.

\compdep{}

We direct our attention first to the Raman peak observed at 128 {\cm} (box i of figure~\ref{introraman}(e), figure~\ref{compdep}(a)). For x = 0, this feature corresponds to an $A_g$ symmetry mode in {\mote}, as demonstrated by our polarization-resolved measurements, DFT calculations \cite{Beams2016}, and other literature observations \cite{Ma2016,Park2015,Chen2016a}. This mode is a single peak for compositions x $\leq$ 0.04, which is consistent with the inversion symmetric {\tp} phase. For compositions x = 0.09 and 0.12, however, the 128 {\cm} mode broadens and is best fit by a pair of Lorentzian functions. Finally, for x $\geq$ 0.29, the 128 {\cm} mode splits into two well-resolved peaks. The separation between these two peaks is presented versus composition in figure~\ref{compdep}(c) and illustrates the appearance and evolution of the doublet, which persists into the $\mbox{x}=1$ (pure {\wte}) case but with a smaller peak separation. Temperature-dependent electrical and XRD measurements have previously shown that {\mote} undergoes a temperature-induced phase transition from the {\tp} to {\td} crystal structure when cooled below 250 K \cite{Hughes1978,Clarke1978}. Recent temperature-dependent Raman measurements in Ref.\cite{Chen2016a} have also demonstrated that the $A_g$ mode at 128 {\cm} in {\mote} splits into a doublet with $A_1$ mode symmetry due to inversion-symmetry breaking upon transitioning into the {\td} phase at low temperatures. Our XRD measurements identify the $0.04 < \mbox{x} < 0.63$ region as two-phase, and therefore we cannot interpret the appearance of the doublet as signifying a phase transition from {\tp} to {\td}. Instead, we attribute the doublet to the breakdown of inversion symmetry in the {\tp}-{\mowte} alloys, which originates not from a {\tp} $\rightarrow$ {\td} phase transition, but instead from the random substitution of Mo atoms with W atoms. Alloying therefore provides a means of destroying inversion symmetry without eliminating the {\tp} phase. From these observations, it is apparent that x $\geq$ 0.29 W concentration is sufficient to drive a breakdown of inversion symmetry and suggests that Weyl physics may be observable even in this two-phase regime \cite{Soluyanov2015,Sun2015}.

We find that other modes also display sensitivity to compositional disorder and the substitution of Mo for W atoms. Box ii of figure~\ref{introraman}(e) isolates {\mote} and {\wte} Raman modes that evolve with changing composition, as well as a mode at 178 {\cm} that is not present in pure {\tp}-{\mote} or {\td}-{\wte}. We summarize the composition-dependent relative intensities for these three peaks in figure~\ref{compdep}(d). The \enquote{{\mote} peak} refers to the feature at 192~{\cm} (black squares) that is present only in Mo-rich compositions (small x) and is assigned as a $B_g$ symmetry mode in {\mote}. The \enquote{{\wte} peak} is the large 212~{\cm} feature (red circles) present only in W-rich compositions (large x) and is assigned as an $A_1$ symmetry mode in {\wte}. Finally, the \enquote{disorder peak} refers to the 178 {\cm} mode unique to the alloys. The polarization dependence of the disorder peak in the x = 0.29 composition can be seen in figure S6 of the Supporting Information. The {\mote} and {\wte} peaks appear to faithfully track the removal and addition of each atomic species, while the disorder mode appears at x = 0.02, peaks at x = 0.33, and disappears at x = 1. The observed frequency agrees with an infrared-active, but Raman-forbidden, $A_u$ mode at 177~{\cm} predicted by our DFT calculations (Table \ref{modetable}). We therefore suggest that the disorder mode originates from an infrared mode that is activated by the loss of translation symmetry in the lattice. The combined effects of lattice disorder and reduced Mo content at large x values drives the mode to reach its maximum intensity at x = 0.33, which also is the point where the ratio of the normalized intensities of the {\mote} peak to the {\wte} peak approach unity. We note that similar activations of infrared modes by alloy disorder have been previously observed, particularly in Ga$_\mathrm{x}$Al$_\mathrm{1-x}$As \cite{Parayanthal1984}.

Given the non-destructive nature and wide-spread use of Raman spectroscopy, it is desirable to determine alloy composition using a Raman-based method. The {\mote} Raman modes present near 260~{\cm} (box iii, figure~\ref{introraman}(f)) provide a potential measure of the alloy composition, which we demonstrate in figures~\ref{compdep}(e,f). We observe a pair of broad Raman modes near 260~{\cm} in {\mote} that are assigned as $A_g$ modes (box iii of figure~\ref{introraman}(f) and figure~\ref{compdep}(e)) and have been seen in prior studies \cite{Ma2016,Beams2016}. By fitting these two peaks in each spectrum to Lorentzians, we can track the peak separation with composition. We find that the separation between these two features increases with increasing x, and that we can use it to estimate global W content in a {\mowte} crystal (figure~\ref{compdep}(f)). Our results indicate that this method will be effective for x $>$ 0.09, and is therefore most appropriate for higher W concentrations.

\scm{}

Finally, we examine the primary {\wte} peak at 212~{\cm} which is broad and asymmetric upon its appearance at $\mbox{x}=0.29$, but sharpens as $\mbox{x}\rightarrow 1$ (figure~\ref{compdep}(b)). We magnify this feature in figure~\ref{scm}(a) for select compositions. The asymmetric lineshape of the 212~{\cm} peak provides valuable information regarding the incorporation of W into the {\mowte} lattice. We find that the asymmetry of this feature and its evolution with x can be well understood in the context of the phonon confinement model, also referred to as the spatial correlation model \cite{Parayanthal1984,Mignuzzi2015}. The phonon confinement model accounts for relaxation of the $\vec{q}=0$ Raman selection rule by multiplying the Lorentzian function, used to represent standard Raman peaks in a pure crystal, with a Gaussian function of the form $\exp(- q^2 L_C^2 /4)$. Thus, the intensity $I$ of Raman peaks in the phonon confinement model is given by \cite{Parayanthal1984}

\begin{equation}
{I(\omega) \propto \int_{BZ} \exp{\left( \frac{-q^{2}L_C^{2}}{4} \right)} \frac{d^{3}q}{{\left[ \omega-\omega(q) \right]}^2 + {(\Gamma_{0}/2 )}^2}}\text{ },
\label{pcm}
\end{equation}

\noindent{}where $\vec{q}$ is in units of $(2 \pi /a,2\pi / b,2 \pi / c)$, $a = 6.3109$ {\AA}, $b = 3.5323$ {\AA}, and $c = 14.4192$ {\AA} are the DFT-relaxed lattice parameters of {\wte}, $\Gamma_0 = 3.80$ {\cm} is the full width at half maximum of the W peak for composition $\mbox{x}=1$, $\omega(q)$ is the dispersion relation which we calculate from DFT and shift to match the experimental value of $\omega(0)$ (figure S1 of the Supporting Information), and $L_C$ is the phonon correlation length. In a pure crystal, $L_C$ is infinite due to the translational symmetry of the lattice and results in plane wave eigenstates. The Gaussian factor in Equation \eqref{pcm}, in this case, is zero for all $\vec{q}$ except the $\Gamma$ point, and therefore the $\vec{q}=0$ Raman selection rule is preserved. However, {\mowte} alloys exhibit potential fluctuations due to the substitutional doping on the transition metal sublattice. The random positioning of the dopant atoms destroys translational symmetry in the crystal, thereby yielding a finite $L_C$ and relaxing the $\vec{q}=0$ Raman selection rule. We fit the 212~{\cm} W peak for x $\geq$ 0.29 in background-subtracted Raman spectra with this model (red lines in figure~\ref{scm}(a)) using experimentally-derived parameters and the DFT calculated phonon dispersion. The extracted phonon correlation length $L_C$ is plotted versus x in figure~\ref{scm}(b), and is found to increase rapidly with x.

\hraman{}

We now comment on the Raman spectra  of {\hp}-{\mowte} ($\mbox{x}=0 \to 0.09$), which are shown in figure~\ref{hraman}(a). The $A_{1g}$ (173~\cm{}), $E_{2g}^1$ (234~\cm{}), and $B_{2g}^1$ (289~\cm{}) modes are visible in all compounds (figure~\ref{hraman}(a)) and exhibit only small changes. The shifts in mode frequency and linewidth for the $A_{1g}$ and $E_{2g}^1$ modes are summarized in figures~\ref{hraman}(b,c). For $\mbox{x}=0.09$, we find that the $A_{1g}$ and $E_{2g}^1$ modes develop asymmetric tails on the low and high energy sides of the peak, respectively. This asymmetry originates from a finite phonon correlation length as discussed previously and we note that the direction of the tail for each mode is consistent with the phonon dispersions of {\hp}-{\mote} \cite{Goldstein}. In addition, we identify a feature appearing at 202~{\cm} for $0.02 \leq \mbox{x} \leq 0.09$ alloys which we assign as a double-resonance Raman mode originating from the scattering of two longitudinal acoustic phonons from the $M$-point or an $E_{1g}(M)$ and a transverse acoustic mode, both also from the $M$ point \cite{Guo2015}. This feature has only been observed in few-layer {\hp}-{\mote} under resonant excitation \cite{Guo2015}, and its appearance in the bulk alloy samples is believed to originate from an enhancement in $\vec{q}\neq 0$ Raman scattering processes by compositional disorder in the lattice.

\section{Conclusion}

We have used XRD, STEM, DFT, and Raman spectroscopy to characterize the different crystal phases spanned by the {\mowte} alloy system. XRD and STEM measurements determined that {\tp}/{\td}-{\mowte} alloys are in the {\tp} phase for $\mbox{x} \leq 0.04$ and the {\td} phase for $\mbox{x} \geq 0.63$. For compositions $0.04 <\mbox{x}<0.63$, {\mowte} exists in a {\tp} + {\td} two-phase mixture. Raman measurements enable the assignment of phonon mode symmetries across the compositional phase space and permit the observation of a new disorder-activated mode unique to {\mowte} alloys. Furthermore, we find that inversion symmetry breaking in the {\tp} phase can occur without transitioning to an orthorhombic configuration by monitoring the splitting of the 128 {\cm} peak. Finally, we find that the asymmetry of the primary {\wte} peak can be captured by the phonon confinement model, which in turn allows for the determination of the phonon correlation length. Our studies of the {\hp} phase show small changes in mode frequencies with x and provide evidence for disorder enhancement of double-resonance Raman scattering processes. These measurements are foundational for future studies seeking to explore the electronic, vibrational, or topological properties of {\mowte} alloys.   

\section{Acknowledgements}
S.M.O., J.J., I.R.S., and P.M.V. acknowledge support from the Office of Naval Research through Grant No. N-00014-15-1-2357, the George Mason University OSCAR Program, and the George Mason University Presidential Scholarship Program. R.B. thanks the National Research Council Research Associateship Programs for its support.  S.K. acknowledges support from the U.S. Department of Commerce, National Institute of Standards and Technology under the financial assistance award 70NANB16H043. A.K.S. is funded by the Professional Research Experience Postdoctoral Fellowship under award No. 70NANB11H012. This research used computational resources provided by the Texas Advanced Computing Center under Contract TG-DMR150006. This work used the Extreme Science and Engineering Discovery Environment (XSEDE), which was supported by National Science Foundation grant number ACI-1053575. A.V.D., S.K., I.K., and A.B. acknowledge the support of Material Genome Initiative funding allocated to NIST.

\end{document}